\documentclass{article}

\usepackage{arxiv}

\usepackage[utf8]{inputenc} 
\usepackage[T1]{fontenc}    
\usepackage{hyperref}       
\usepackage{url}            
\usepackage{booktabs}       
\usepackage{amsfonts}       
\usepackage{nicefrac}       
\usepackage{microtype}      
\usepackage{lipsum}
\usepackage{amsmath}
\usepackage{graphicx,psfrag,epsf}
\usepackage{bm}
\usepackage{color}
\usepackage{mathrsfs}
\usepackage[T1]{fontenc}
\newcommand{\choosefont}[1]{\fontfamily{#1}\selectfont} 

\title{ Concordance Rate of a Four-Quadrant Plot \\for Repeated Measurements}

\author{
  Mayu Hiraishi \\
  Clinical Study Support Center\\
  Wakayama Medical University Hospital\\
   \And
 Kensuke Tanioka \\
   Faculty of Life and Medical Sciences\\
  Doshisha University\\
   \AND
   Toshio Shimokawa \\
    Department of Medical Data Science \\
    Wakayama Medical University
}

\begin{document}
\maketitle

\begin{abstract}
Before new clinical measurement methods are implemented in clinical practice, it must be confirmed whether their results are equivalent to those of existing methods. The agreement of the trend between these methods is evaluated using the four-quadrant plot, which describes the trend of change in each difference of the two measurement methods’ values in sequential time points, and the plot's
concordance rate, which is calculated using the sum of data points  in the four-quadrant plot that agree with this trend divided by the number of all accepted data points.
However, the conventional concordance rate does not consider the covariance between the data on individual subjects, which may affect its proper evaluation.
Therefore, we proposed a new concordance rate calculated by each individual according to the number of agreement. Moreover, this proposed method can set a parameter that the minimum concordant number between two measurement techniques. The parameter can provide a more detailed interpretation of the degree of agreement.
A numerical simulation conducted with several factors indicated that the proposed method resulted in a more accurate evaluation. We also showed a real data and compared the proposed method with the conventional approach. Then, we concluded the discussion with the implementation in clinical studies.
\end{abstract}


\section{Introduction}

New clinical measurements and new technologies such as cardiac output (CO) monitoring continue to be introduced, and it must be verified whether the results of the new testing measurement methods are equivalent to those of the standard measurement methods before implementing them in clinical practice. For example, an improved cardiac index (CI) tracking device was compared to a traditional method for CI by transpulmonary thermodilution to assess the reliability for accurately measuring changes in norepinephrine dose during operations (Monnet {\it{et al}}., 2012).
In the study of Cox {\it{et al}}. (2017), the bioimpedance electrical cardiometry, another experimental measurement device of CI, examined with the continuous pulmonary artery thermoregulatory catheterization as the gold standard by conducting before, during, and after cardiac surgery. 

Various statistical methods have been proposed to assess the equivalence of the new testing measurement methods with the gold standards (e.g., Carstensen, 2010;
Choundhary and Nagaraja, 2005;
Choudhary and Nagaraja, 2017).
In Altman and Bland (1983), Bland and Altman (1986) and 
Bland and Altman (1996),
the Bland-Altman analysis has been proposed
to evaluate the accuracy of a new clinical test based on its difference from a gold standard measurement values and on the mean of the two tests values. 
In addition, a method for calculating the sample size when conducting the Bland-Altman analysis during clinical trials has been proposed by Shieh (2019). The Bland-Altman analysis has also expanded to cases of repeated measurement (e.g., Bland and Altman, 2007; Bartko, 1976; Zou, 2013), which have been used in clinical studies. Asamoto {\it{et al}}. (2017) used the Bland-Altman analysis to evaluate the equivalence of the accuracy in the less invasive continuous CO monitor during two different surgeries. However, the Bland-Altman plot can not describe the trending ability between the two compared measurements, because the Bland-Altman analysis does not consider the order of the observed data. 
Thus, to evaluate the trending ability, the researchers also showed the four-quadrant plot for drawing the changes of the measurement results and calculated the concordance rate. 
In fact, in these equivalence comparative clinical trials, the four-quadrant plot and the concordance rate are often used along with the Bland-Altman analysis. 

As the assessment based on the degree of trending of the CO changes at each time point, the use of the four-quadrant plot and concordance rate has been proposed (Perrino {\it{et al}}., 1994; Perrino {\it{et al}}., 1998). 
The four-quadrant plot and four-quadrant concordance analysis are often employed with the Bland-Altman analysis when evaluating the equivalence of the two measurement methods (e.g. Monnet {\it{et al}}., 2012). 
The four-quadrant plot and concordance rate focus on the trending ability between each difference of two testing values, while  Bland-Altman analysis assesses the accuracy and the precision of values of two measurement methods. In a four-quadrant plot, pairs of each difference of two testing values at sequential time points are plotted. For example, a plot draws with the value at the second time point minus the value measured at the first time point which are measured by the gold standard on the horizontal axis, and the difference value between the same time points measured by the experimental method on the vertical axis. 

The evaluation of the four-quadrant plot is based on whether the trends regarding each difference between the new experimental measurement and the gold standard are concordant. 
When the trends between the two measurements increase or decrease together, those points are regarded as being in agreement (Saugel {\it{et al}}., 2015). 
Here, the small difference values do not count for the concordance rate by introducing the ``exclusion zone".

Concordance rate in a four-quadrant plot is calculated by the ratio of the number of agreements to all data points. However, this conventional concordance rate does not consider covariance within an individual, even though, in general, one subject is measured multiple times in clinical practice.
In the case when the covariance within an individual is high, this may lead to incorrect results in the calculation without considering the covariance.
However, concordance rate for the four-quadrant plot has not been expanded for repeated measurement, unlike the Bland-Altman analysis.

Thus, our study proposes a new concordance rate for the four-quadrant plot based on multivariate normal distribution in order to take into account the individual subjects.
This new method can be applied to any number of repeated measurement.
Specifically, the proposed concordance rate is formulated as conditional probabilities of the agreement given the event in which no data points within individual fall into the exclusion zone.  
In this study, we examine the case of three time points in numerical simulation.

The proposed method also has a parameter to set the minimum concordant number $m$ between two measurement methods regarded as being in ``agreement". The concordance number is counted based on how many times an ``agreement" out of the number of differences of measurement values $T$. This parameter is the least number that the trending of  two clinical measurement methods can be assessed in calculating concordance rate. For instance, when the parameter $m$ is $3$ and $T$ is $5$, the concordance rate evaluates the case of more than $3$ agreements out of $5$ times.
This parameter allows analysts setting from a clinical perspective.
In general, $T = m$ and the high probability of the concordance are ideal, but the parameter can provide a more detailed interpretation of the degree of agreement by adjusting the parameter $m$. \

Accordingly, this study first proposes the new concordance rate for the four-quadrant plot in a general framework and then takes the case of the calculation at three time points as an example. In detail, the remainder of this paper is organized as follows;
Section 2 explains the general concordance rate for the four-quadrant plot. In Section 3, we introduce the new proposed concordance rate
and present the case wherein the maximum number of agreements is two.
Then, Section 4 presents the application of the proposed method to simulations and its result. Section 5 describes the results of the application to a real example. We conclude this paper in Section 6.

\section{Concordance Rate}
\label{sec:concordance}

This section explains the ways to draw the four-quadrant plot and calculate the concordance rate by using the conventional method. The assessment method for the trending agreement of two testing values using the four-quadrant plot was first proposed by Perrino, {\it{et al}}. (1994). The four-quadrant plot uses each pair of differences between the values measured by the two clinical methods being compared. Point $x^{*}_{it^{*}} (i=1,2,\cdots,n;\quad t^{*}=1,2,\cdots,(T+1))$ indicates as the value of a gold standard for individual subject $i$ at time $t^*$th, and $y^{*}_{it^{*}} (i=1,2,\cdots,n;\quad t^{*}=1,2,\cdots,(T+1))$ is the value of the experimental technique. Then, the $t$th difference of the values measured by the gold standard is
\begin{align*}
x_{it} = x^{*}_{i(t+1)} - x^{*}_{it} \quad (t=1,2,\cdots,T),
\end{align*}
and the $t$th difference of the values measured by the experimental technique is
\begin{align*}
y_{it} = y^{*}_{i(t+1)} - y^{*}_{it} (t=1,2,\cdots,T).
\end{align*}

\begin{figure}[htbp]
\begin{center}
\includegraphics[scale=0.5]{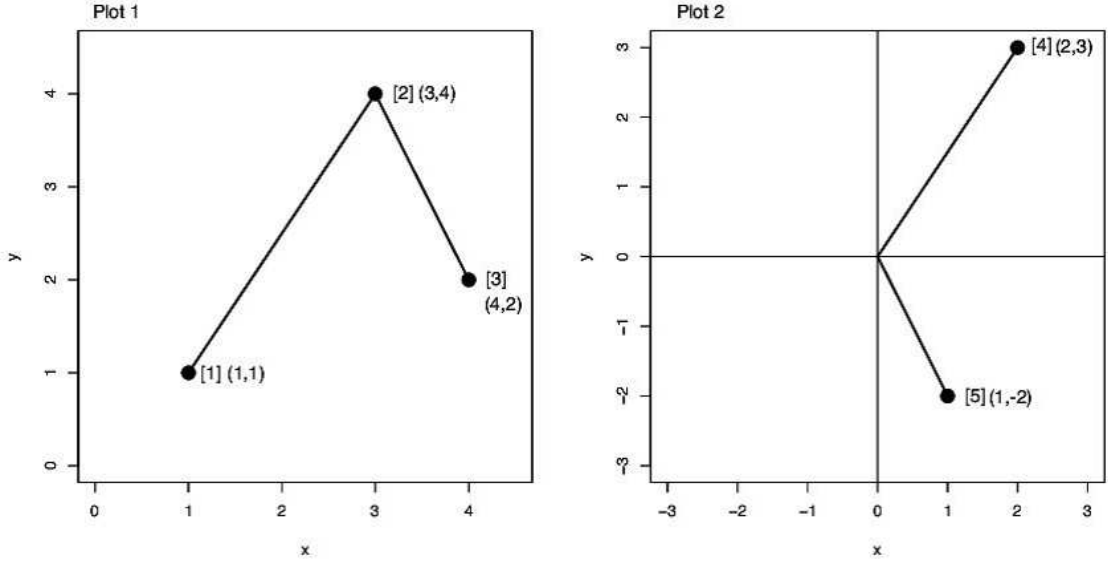}
\caption{Plots for the step of drawing the four-quadrant plot. The horizontal axis denotes $x$, and the vertical axis denotes $y$. Plot 1: Data plotted for three pairs of values on Cartesian coordinates. Plot 2: Four-quadrant plot of the data in Plot 1.}
\label{4q step}
\end{center}
\end{figure}

Plot 1 in Figure \ref{4q step} shows an example of treatment values in a time sequence that compares two tests for one subject. Focusing on the first two data points in Plot 1, the difference between [2] and [1] can be described as [4] of the four-quadrant plot in Plot 2. At this time, both $x$ and $y$ increase, which indicates that the direction of change in $x$ and $y$ is the same. A point such as [4] plotted in the upper-right of the four-quadrant plot can be evaluated as being in ``agreement." In contrast, the difference between [3] and [2] is plotted as [5] in the lower-right of Plot 2. In this case, $x$ increases but $y$ decreases, which means that the trend of $x$ and $y$ is recognized as being in ``disagreement." Similarly, if the difference in both $x$ and $y$ is negative, as plotted in the lower-left, the change is also in ``agreement," while the data points in the upper-left can be assessed as being in ``disagreement."

\begin{figure}[htbp]
\begin{center}
\includegraphics[scale=0.60]{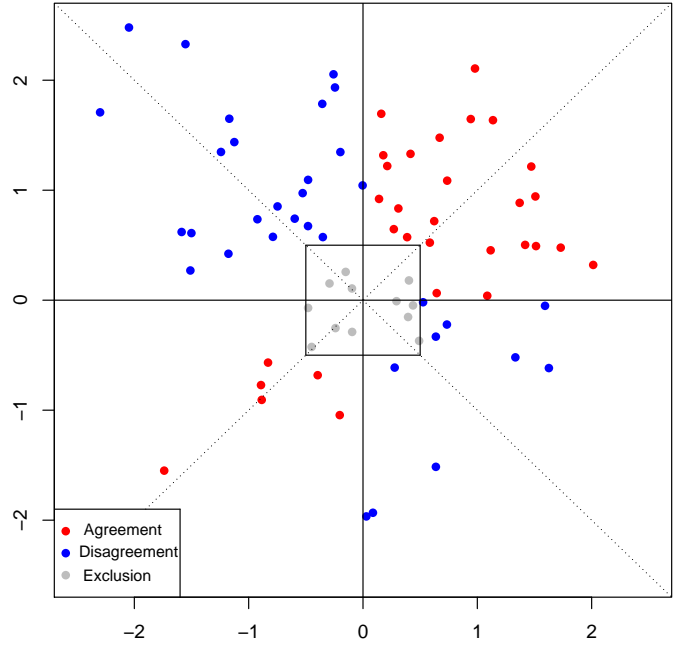}
\caption{Four-quadrant plot with artificial example data.}
\label{4q plot}
\end{center}
\end{figure}

Figure \ref{4q plot} is a four-quadrant plot with artificial example data. In the figure, the red points in the upper-right and lower-left sections are counted as being in ``agreement." The blue dots, on the other hand, signify ``disagreement." When the difference value of the experimental technique is equal to that of the gold standard, the data dot is on the $45^\circ$ lines (dotted lines in Figure \ref{4q plot}). 

The concordance rate is calculated based on the idea above. The conventional concordance rate (CCR) is defined as follows:
\begin{align}
{\rm CCR}(a) =
\frac{ \# {\rm SA} - \# {\rm AEz}(a) }{ nT - \# {\rm Ez} (a) },
\end{align}
where
\begin{align*}
{\rm SA} =& \{(x_{it}, y_{it})| \; {\big(} (x_{it}\geq0,\;y_{it}\geq0)\ \cup\ (x_{it}<0,\;y_{it}<0){\big)}, \\
&i=1,2,\cdots,n;\;t=1,2,\cdots,T \}, \\
{\rm AEz}(a) =& \{(x_{it}, y_{it})| \; {\big(} (0\leq x_{it}\leq a,\;0\leq y_{it}\leq a)\  \cup\ (-a< x_{it}<0,\;-a<y_{it}<0){\big)} \\
&i=1,2,\cdots,n;\;t=1,2,\cdots,T \}, \quad {\rm and } \\
{\rm Ez}(a)=&\{(x_{it},y_{it})|\;-a\leq x_{it},x_{it} \leq a,\;-a \leq y_{it},y_{it} \leq a,\; t=1,2,\cdots,T\}.\\
\end{align*}
{\rm SA} is the set of ``agreement" pairs of each difference between the values of the gold standard and experimental technique. ${\rm Ez}(a)$ is the set of pairs plotted in the exclusion zone. In the four-quadrant plot, the exclusion zone (middle square in Figure \ref{4q plot}) is usually placed to remove data plots close  to the origin of the plot, because it is difficult to determine whether such small values have occurred due to the examination or mechanical errors (e.g.,Critchley {\it{et al}}., 2010). The gray points plotted in the exclusion zone in Figure \ref{4q plot} are excluded when calculating the concordance rate. The range of the exclusion zone depends on $a$, which is set from a clinical point of view (e.g.,Saugel {\it{et al}}., 2015). ${\rm AEz}(a)$ is the set of the ``agreement" pairs in the exclusion zone.  \# signifies the cardinality of a set. The concordance rate in Eq. (1) is the ratio between the number of data points in the ``agreement" sections except exclusion zone with all data points that fall outside the exclusion zone.

This conventional concordance rate simply counts the number of data points that show the same trend of change. However, multiple measurements are generally taken for a single patient in a clinical setting. Individual tendencies may influence the measurement results for a single subject. Therefore, individuals must be considered to calculate a more precise concordance rate. 

\section{Concordance Rate for the Four-quadrant Plot}
\label{sec:proposal}

\subsection{General framework of the proposed concordance rate}

The proposed concordance rate evaluates the equivalence between the experimental technique and the gold standard through calculation that considers the individual subjects. This proposed method includes the exclusion zone as well, and is defined as the conditional probability, which corresponds to the event falling out of the exclusion zone in all time points. We estimate the parameters of the population with all the data.

The approach for calculation of the proposed method starts with the four-quadrant plot per point $t$. First, the quadrant sections are named $A_t$ to $D_t$. The sample space where the $t$th value falls in each section can be described in four ways:
\begin{align*}
    A_{t}=&\{\omega |\; X_{t} (\omega) \geq 0, Y_{t} (\omega) \geq 0\}, \\
    B_{t}=&\{\omega |\; X_{t} (\omega) < 0, Y_{t} (\omega) < 0\}, \\
    C_{t}=&\{\omega |\; X_{t} (\omega) < 0, Y_{t} (\omega) \geq 0\}, \quad {\rm and} \\
    D_{t}=&\{\omega |\; X_{t} (\omega) \geq 0, Y_{t} (\omega) < 0\} \quad (t = 1, 2, \cdots, T).
\end{align*}
Here, $X_t$ and $Y_t$ are random variables of each difference of the values of the gold standard and experimental techniques, respectively. $X_t$ and $Y_t$ correspond to $x_{it}$ and $y_{it}$, respectively. ${\rm X} = (X_1, X_2, \cdots, X_T)$ and ${\rm Y} = (Y_1, Y_2, \cdots, Y_T)$ are assumed to be distributed from multivariate normal distributions. 
$A_t$ in the upper-right and $B_t$ in the lower-left quadrants of the four-quadrant plot (Figure \ref{4q plot}) correspond with ``agreement," whereas $C_t$ in the upper-left and $D_t$ in the lower-right quadrants are in ``disagreement."

Here, the family of sets is defined as follows:
\begin{align*}
\mathscr{W}_t = \{A_t \cup B_t, C_t \cup D_t\} 
\quad (t = 1, 2, \cdots, T).
\end{align*}
Then, exclusion zone at the $t$th time is 
\begin{align*}
    {\rm Ez_t}(a)=&\{\omega|\;-a \leq X_t(\omega) \leq a, -a \leq Y_t(\omega) \leq a\} \quad (t = 1, 2, \cdots, T).
\end{align*}
${\rm Ez}(a)$ is also divided into four-quadrant sections:
\begin{align*}
    {\rm EzA_t}(a)=&\{\omega|\;0 \leq X_t(\omega) \leq a, 0 \leq Y_t(\omega) \leq a\}, \\
    {\rm EzB_t}(a)=&\{\omega|\;-a \leq X_t(\omega) \leq 0, -a \leq Y_t(\omega) \leq 0\}, \\
    {\rm EzC_t}(a)=&\{\omega|\;-a \leq X_t(\omega) \leq 0, 0 \leq Y_t(\omega) \leq a\}, \\
    {\rm EzD_t}(a)=&\{\omega|\;0 \leq X_t(\omega) \leq a, -a \leq Y_t(\omega) \leq 0\}\quad (t = 1, 2, \cdots, T).
\end{align*}

The assets of the random variables in $A_t, B_t, C_t$, and $D_t$, except the exclusion zone, are defined as follows:
\begin{align*}
A_t^{\dagger} =& A_t \cap {\rm EzrA_t}(a)^c,  \\
B_t^{\dagger} =& B_t \cap {\rm EzrB_t}(a)^c,  \\
C_t^{\dagger} =& C_t \cap {\rm EzrC_t}(a)^c,  \quad {\rm and}\\
D_t^{\dagger} =& D_t \cap {\rm EzrD_t}(a)^c ,
\end{align*}
where $Z^c$ is the complement of arbitrary set $Z$.\
$A_t^{\dagger}$ and $B_t^{\dagger}$ are the events of ``agreement" that do not fall into the exclusion zone, whereas $C_t^{\dagger}$ and $D_t^{\dagger}$ are the events of ``disagreement" out of the exclusion zone.

The proposed concordance rate
is calculated in the condition when all pairs of $(X_t,Y_t)$ are not in the exclusion zone. This means that all data of one subject are excluded from the calculation if any pair of data points for that subject drops to the exclusion zone at least once. This can be described as
\begin{align*}
{\rm NEz(a)} = \Big \{ \omega|\;  \forall t \; (t = 1,2, \cdots, T); \; \omega \notin {\rm Ez}_t(a)\Big \}. 
\end{align*} 
Here, the two clinical testing methods are regarded as equivalent if $X_t$ and $Y_t$ show the same direction of trends more than $m$ times out of $T$ times per subject. 
Concordance rate in the agreement times more than the setting number in $m$ is calculated. $m$ is determined from a clinical perspective. $T$ is the number of differences of measurement values.
Given this idea, we propose the new concordance rate, wherein the probability of ``agreement" of more than $m$ times in $T$ is defined as follows: 
\begin{align}
P\Big[ \bigcup_{t = m}^T
H_t | {\rm NEz}(a) 
 \Big] \nonumber \
=& \frac{
P\Big[ 
(\bigcup_{t = m}^T H_t) \cap {\rm NEz}(a)
 \Big] }
{ P\Big [  {\rm NEz}(a) \Big] }\\
=& \frac{
\sum_{t = m}^T
P\Big[ 
H_t \cap {\rm NEz}(a)
 \Big] }
{ 1 - P\Big [ \bigcup_{s=1}^T {\rm Ez_s}(a) \Big] },
\label{condition1}
\end{align}
where
\begin{align}
H_t = \Big \{ \omega |\;  (W_1(\omega), W_2(\omega), \cdots, W_T(\omega)) \in \prod_{s=1}^T \mathscr{W}_s, 
\sum_{s=1}^T I(W_s(\omega) = A_s(\omega) \cup B_s(\omega))  = t 
\Big \}.  
\label{condition2}
\end{align}

$H_t$ in Eq. (\ref{condition1}) is the subset of the sample space wherein the trend between $X$ and $Y$ agrees $t$ times. $I$ is the indicator function in the condition wherein the $s$th data fall in $A^{\dagger}$ or $B^{\dagger}$. $\prod_{s=1}^T \mathscr{W}_s$ in Eq. (\ref{condition2}) indicates the product.

\subsection{Example of the proposal index, T = 2}

Next, we explain the proposed concordance rate in the case of $m = 1$ and $T = 2$, that is, at three points in time. The probability can be calculated as follows:
\begin{align}
P\Big[ \Big( \bigcup_{t = 1}^2
H_t \Big) \cap {\rm NEz}(a) 
 \Big] \
= \frac{
\sum_{t = 1}^2
P\Big[ 
H_t \cap {\rm NEz}(a)
 \Big] }
{ 1 - P\Big [ \bigcup_{s=1}^2 {\rm Ez}_s(a) \Big] }. \label{condition3}
\end{align}

We apply the definition at $T =2$ to a four-quadrant plot. There are three patterns in the case of $T =2$: agreement in $t=1$, agreement in $t=2$, and agreements in $t=1$ and $t=2$. The probability of the numerator in the definition formula is 
\begin{align}
    P[H_1 \cap {\rm NEz}(a)] =& P[(A_1^{\dagger} \cup B_1^{\dagger}) \cap (C_2^{\dagger} \cup D_2^{\dagger})] 
+ P[(C_1^{\dagger} \cup D_1^{\dagger}) \cap (A_2^{\dagger} \cup B_2^{\dagger})]\label{eq2}  \\
    P[H_2 \cap {\rm NEz}(a)] =& P[(A_1^{\dagger} \cup B_1^{\dagger}) \cap (A_2^{\dagger} \cup B_2^{\dagger})]. \label{eq3} 
\end{align}

To describe each case, the range wherein the data point enters into each quadrant of the plot is set as $F = \{ [0, \infty]^T, \; [-\infty, 0]^T \}$, and the range of the exclusion zone is $E = \{ [0, a]^T, \; [-a, 0]^T \}$. 
Vectors to describe the range for the probability calculations are as follows:
\begin{align*}
{\bf v_1} =
\left[
\begin{array}{c}
v_{11} \\
v_{21} \\
\end{array}
\right] , \quad
{\bf v_2} =
\left[
\begin{array}{c}
v_{12} \\
v_{22} \\
\end{array}
\right], \quad 
{\bf z_1} =
\left[
\begin{array}{c}
z_{11} \\
z_{21} \\
\end{array}
\right], \quad 
{\bf z_2} =
\left[
\begin{array}{c}
z_{12} \\
z_{22} \\
\end{array}
\right].
\end{align*}
The first term of Eq. (\ref{eq2}) means  the probability with which the trend of $X_1$ and $Y_1$ is in agreement, whereas that of $X_2$ and $Y_2$ is not. This can also be  expressed as 
\begin{align*}
&P\Big[ (A_1^{\dagger} \cup B_1^{\dagger}) \cap  (C_2^{\dagger} \cup D_2^{\dagger})  \Big]  \nonumber \\
=& 
\sum_{\substack{{\bf v_1} = {\bf z_1}, {\bf v_2} \neq {\bf z_2} \\
{\bf v_1}, {\bf v_2}, {\bf z_1}, {\bf z_2} \in F  }}
P(v_{11} < X_1 < v_{21},\; v_{12} < X_2 < v_{22},\;  
z_{11} < Y_1 < z_{21},\; z_{12} < Y_2 < z_{22}  ) \nonumber  \\
&+
\sum_{\substack{{\bf v_1} = {\bf z_1}, {\bf v_2} \neq {\bf z_2} \\ 
{\bf v_1},{\bf v_2}, {\bf z_1}, {\bf z_2} \in E  }}
P(v_{11} < X_1 < v_{21},\; v_{12} < X_2 < v_{22},\;  
z_{11} < Y_1 < z_{21},\; z_{12} < Y_2 < z_{22}  ) \nonumber  \\
&-
\sum_{\substack{{\bf v_1} = {\bf z_1}, {\bf v_2} \neq {\bf z_2} \\ 
{\bf v_1}, {\bf z_1} \in F, \;  {\bf v_2}, {\bf z_2} \in E  }}
P(v_{11} < X_1 < v_{21},\; v_{12} < X_2 < v_{22},\;  
z_{11} < Y_1 < z_{21},\; z_{12} < Y_2 < z_{22}  ) \nonumber  \\
&-
\sum_{\substack{{\bf v_1} = {\bf z_1}, {\bf v_2} \neq {\bf z_2} \\ 
{\bf v_1}, {\bf z_1} \in E, \;  {\bf v_2}, {\bf z_2} \in F }}
P(v_{11} < X_1 < v_{21},\; v_{12} < X_2 < v_{22},\;  
z_{11} < Y_1 < z_{21},\; z_{12} < Y_2 < z_{22}  ).
\end{align*}
Then, the second term of Eq. (\ref{eq2}) is the probability when the trend of $X_1$ and $Y_1$ is in disagreement, but that of $X_2$ and $Y_2$ is in agreement. This can be rewritten similarly as
\begin{align*}
&P\Big[ (C_1^{\dagger} \cup D_1^{\dagger}) \cap  (A_2^{\dagger} \cup B_2^{\dagger})  \Big]  \nonumber \\
=& 
\sum_{\substack{\bf{v}_1 \neq {\bf z_1}, {\bf v_2} = {\bf z_2} \\ 
{\bf v_1}, {\bf v_2}, {\bf z_1}, {\bf z_2} \in F  }}
P(v_{11} < X_1 < v_{21},\; v_{12} < X_2 < v_{22},\;  
z_{11} < Y_1 < z_{21},\; z_{12} < Y_2 < z_{22}  ) \nonumber  \\
&+
\sum_{\substack{{\bf v_1} \neq {\bf z_1}, {\bf v_2} = {\bf z_2} \\
{\bf v_1}, {\bf v_2}, {\bf z_1}, {\bf z_2} \in E  }}
P(v_{11} < X_1 < v_{21},\; v_{12} < X_2 < v_{22},\;  
z_{11} < Y_1 < z_{21},\; z_{12} < Y_2 < z_{22}  ) \nonumber  \\
&-
\sum_{\substack{{\bf v_1} \neq {\bf z_1}, {\bf v_2} = {\bf z_2} \\
{\bf v_1}, {\bf z_1} \in F, \;  {\bf v_2}, {\bf z_2} \in E  }}
P(v_{11} < X_1 < v_{21},\; v_{12} < X_2 < v_{22},\;  
z_{11} < Y_1 < z_{21},\; z_{12} < Y_2 < z_{22}  ) \nonumber  \\
&-
\sum_{\substack{{\bf v_1} \neq {\bf z_1}, {\bf v_2} = {\bf z_2} \\
{\bf v_1}, {\bf z_1} \in E, \;  {\bf v_2}, {\bf z_2} \in F  }}
P(v_{11} < X_1 < v_{21},\; v_{12} < X_2 < v_{22},\;  
z_{11} < Y_1 < z_{21},\; z_{12} < Y_2 < z_{22}  ).
\end{align*}
Eq. (\ref{eq3})
is the probability that the trends of $X_1$ and $Y_1$ and of $X_2$ and $Y_2$ are both concordant:
\begin{align*}
&P\Big[ (A_1^{\dagger} \cup B_1^{\dagger}) \cap  (A_2^{\dagger} \cup B_2^{\dagger})  \Big]  \nonumber \\
=& 
\sum_{\substack{{\bf v_1} = {\bf z_1}, {\bf v_2} = {\bf z_2} \\ 
{\bf v_1}, {\bf v_2}, {\bf z_1}, {\bf z_2} \in F  }}
P(v_{11} < X_1 < v_{21},\; v_{12} < X_2 < v_{22},\;  
z_{11} < Y_1 < z_{21},\; z_{12} < Y_2 < z_{22}  ) \nonumber  \\
&+
\sum_{\substack{{\bf v_1} = {\bf z_1}, {\bf v_2} = {\bf z_2} \\
{\bf v_1}, {\bf v_2}, {\bf z_1}, {\bf z_2} \in E  }}
P(v_{11} < X_1 < v_{21},\; v_{12} < X_2 < v_{22},\;  
z_{11} < Y_1 < z_{21},\; z_{12} < Y_2 < z_{22}  ) \nonumber  \\
&-
\sum_{\substack{{\bf v_1} = {\bf z_1}, {\bf v_2} = {\bf z_2} \\ 
{\bf v_1}, {\bf z_1} \in F, \;  {\bf v_2}, {\bf z_2} \in E  }}
P(v_{11} < X_1 < v_{21},\; v_{12} < X_2 < v_{22},\;  
z_{11} < Y_1 < z_{21},\; z_{12} < Y_2 < z_{22}  ) \nonumber  \\
&-
\sum_{\substack{{\bf v_1} = {\bf z_1}, {\bf v_2} = {\bf z_2} \\ 
{\bf v_1}, {\bf z_1} \in E, \;  {\bf v_2}, {\bf z_2} \in F  }}
P(v_{11} < X_1 < v_{21},\; v_{12} < X_2 < v_{22},\;  
z_{11} < Y_1 < z_{21},\; z_{12} < Y_2 < z_{22}  ).
\end{align*}

Finally, the probability of the denominator in $T =2$ is
\begin{align*}
&1 - P\Big [ \bigcup_{s=1}^2 {\rm Ez}_s(a) \Big]\\
=&
1- P(-a < X_1 < a. -\infty < X_2 <\infty,\;-a < Y_1 <a, -a < Y_2 <a)\\
&-
P(-\infty < X_1 < \infty, -a < X_2 <a,\;-\infty < Y_1 < \infty, -a < Y_2 < a)\\
&+
P(-a < X_1 < a, -a < X_2 <a,\;-a < Y_1 < a, -a < Y_2 < a).
\end{align*}

In the proposed concordance rate, we assume that all random variables are distributed from multivariate normal distribution.
Therefore, we must estimate the mean vectors and covariance matrices to calculate the concordance rate. The method of estimating these parameters is described next.

\subsection{Estimation}\label{sec3}

First, we define 
%
$Z = (X_1,\; \cdots,\; X_T,\; Y_1,\;  \cdots,\; Y_T) 
= (Z_1,\; \cdots,\; Z_{T}, Z_{T+1},\; \cdots,\; Z_{T+T})$.  
%
Since the proposed method assumes that $Z$ are distributed from $T+T$-dimensional normal distributions, it is necessary to estimate the $T+T$-dimensional mean vector and variance covariance matrix to calculate the concordance rate.
The estimated mean vector in the proposed approach is 
%
$\bar{\bm z}=(\bar{x}_1,\;\cdots,\;\bar{x}_T,\; \bar{y}_1,\;\cdots,\;\bar{y}_T )^T$ 
%
, where $\bar{x}_t$ and $\bar{y}_t$ are the mean of the $t$th value of gold standard and experimental technique, respectively. 
The covariance matrix based on the differences between the times is ${\bm{S}} = (s_{tk})\quad (t,k=1,2,\cdots,T+T)$, where $s_{tk}$ is the covariance between $t$ and $k$. 
By using these estimator, the proposed concordance rate in Eq (\ref{condition1}),
defined as the conditional probability, can be calculated.

\section{Numerical Simulation}
\label{sec:4}
In this section, we describe the simulation design, and present the simulation results. We conducted a simulation and set the two types of evaluation for the simulation. 
First, we examined how close the concordance rates calculated with the conventional methods and the proposed approach were to the result of the true concordance rate. The assessment of each concordance rate was expressed as the difference from the true concordance rate. The results of the proposed method can not be simply compared with CCR$(a)$, since CCR$(a)$ does not consider the repeated measurement. In order to compare with the conventional concordance rate, control1 and control2 were adjusted to allow repeated measurements, which details in factor 7.\
Secondly, to compare the diagnosability of the proposed method with CCR$(a)$, we calculated ROC curves and Area Under the Curve (AUC)  (e.g. Pepe, 2003). The second evaluation is based on AUC. 
In this simulation, we used 
{\choosefont{pcr} RStudio Version 1.1.453.}
%

\subsection{Simulation design}

We set $T=2$, and the data generation procedure is as follows:
\begin{align*}
    \bf{Z} \sim  N (\bf{\mu}_{z}, \Sigma_{z})
\end{align*}
where ${\bf Z} = (X_1,X_2,Y_1,Y_2) ^{T}$ .
$X_t$  is the difference in the measurement values of the gold standard between the $t$th and $(t+1)$th times $(t=1,2,3)$, and $Y_t$ is that of experimental technique. 

In addition,  
\begin{align*}
    \bm{\mu}_{Z} = \left[ 
\begin{array}{c}
\bm{\mu}_{X}\\
\bm{\mu}_{Y}\\
\end{array} 
\right]
,
\quad
    \bm{\Sigma}_{Z} = \left[ 
\begin{array}{cc}
\bm{\Sigma}_{X} & \bm{\Sigma}_{XY} \\
\bm{\Sigma}_{XY} & \bm{\Sigma}_{Y}\\
\end{array} 
\right],
\end{align*} 
where $\bf{\mu}_{X} = (\mu_{x1},\mu_{x2})^{T}$   and $\bf{\mu}_{Y} = (\mu_{y1},\mu_{y2})^{T}$ are the mean vectors of the gold standard and experimental technique, and $\bf{\Sigma}_{X}$ and $\bf{\Sigma}_{Y}$ are the covariance matrices, respectively.\\
Here, 

\begin{align*}
   \bm{\Sigma}_{X} = \left[ 
\begin{array}{cc}
{\sigma}_{x1} & {\rho} \\
{\rho} & {\sigma}_{x2}\\
\end{array} 
\right],
\quad
   \bm{\Sigma}_{Y} = \left[ 
\begin{array}{cc}
{\sigma}_{y1} & {\rho} \\
{\rho} & {\sigma}_{y2}\\
\end{array} 
\right], \ {\rm and} \ 
\quad
   \bm{\Sigma}_{XY} = \left[ 
\begin{array}{cc}
{\rho}_{XY} & {\rho}_{XY} \\
{\rho}_{XY} & {\rho}_{XY}\\
\end{array} 
\right].
\end{align*} 

we set $\sigma_{x1} = \sigma_{x2} = \sigma_{y1} = \sigma_{y2} = 1$.

Factors set in the simulation are presented in Table \ref{T1}. 
The number of total patterns is $30\times3\times2\times2\times2\times2\times3=4320$. For each pattern, corresponding artificial data are generated 100 times and we evaluate the results.
The levels of the seven factors are set as follows. 

\

\noindent
{\bf Factor 1: Means}

The mean is of 30 patterns, as shown in Table \ref{T2}. The setting depends on the combination of the magnitude of the mean value and the direction of change in $x$ and $y$.

\

\noindent
{\bf Factor 2: Covariance between the difference values within each measurement method }

The corvariance within each measurement method of the difference values, ${\rho} $, is set as $0$, $1/3$ and $2/3$ in both $X$ and $Y$.

\

\noindent
{\bf Factor 3: Covariance between $X$ and $Y$}

${\rho}_{XY} = 0$ and $1/3$. 

\

\noindent
{\bf Factor 4: Number of agreements}

Factor 4 is the number of trending agreements between $X$ and $Y$. We set two different situations: (1) agreement more than once in $T=2$, and (2) agreement at both time points.

\

\noindent
{\bf Factor 5: Exclusion zone}

$a$ of the exclusion zone ${\rm Ez}(a)$ is set as 0.5 and 1.0. 

\

\noindent
{\bf Factor 6: Number of subjects}

The number of subjects is set as 15 and 40.

\

\noindent
{\bf Factor 7: Methods}

We calculate the concordance rate by four methods. Control1, control2, and the proposed method are used in the first evaluation, and CCR, Control1, control2, and the proposed method are used in the second evaluation. We denote the proposed concordance rate as ``proposal".

Control1, based on binomial distribution, is calculated as follows:
\begin{align*}
    \sum_{s=m}^{2} {}_{2}C_{s} p^s(1-p)^{(2-s)},
\end{align*} 
where 
\begin{align*}
    p = \frac{k_1+k_2}{n_1^{\dagger}+n_2^{\dagger}}.
\end{align*} 
${k_t}$ ($t = 1,2$) is the number of data that show the same trend between $X_t$ and $Y_t$ out of the exclusion zone. $n_t^{\dagger}$ is the number of subjects whose data points fall out of the exclusion zone.
The concordance rate in control2 is calculated by the probability at each number of agreement:
twice in two time points is
$p_1p_2$,
and
once in two time points 
$p_1(1-p_2)+(1-p_1)p_2$,
%

where 
\begin{align*}
    p_t = \frac{k_t}{n_t^{\dagger}}\quad (t=1,2).
\end{align*} \
Subjects whose difference value fall in the exclusion zone of the four-quadrant plot even once are excluded from the calculation of the concordance rate in both control1 and control2 as same manner of the proposed method.

\

The first evaluation index for the simulation result is the absolute values of the difference between the concordance rate based on each estimated parameters and the concordance rate computed with the true mean vector $\bm{\mu}_{Z}$ and with true covariance matrix $\bm{\Sigma}_{Z}$. We set  the evaluation to  deserve as better assessment if the absolute values  of the difference  between the true value  and the estimated values are smaller among all concordance rate approaches.

For the second evaluation index, we label to each pattern of means in Table \ref{T1}. If $\bf{\mu}_{X}$ and $\bf{\mu}_{Y}$ are concordant both two times, we mark the corresponding mean pattern as ``$\circ$", and the rest as ``$\times$". Then, the 4320 $\times$ 100 data in total have this label. ROC and AUC (e.g. Pepe, 2003) are calculated by the label and the results of concordance rates in each method, and we compare these results of AUC among the proposal method, CCR$(a)$, control1 and control2.

\begin{table}
\begin{center}
\caption{Factors of the simulation design} \label{T1}
\begin{tabular}{llll} 
\hline
Factor No. & Factor name & 
levels \\ \hline
Factor 1 & Means & 30 \\ 
Factor 2 & Covariance between the difference values within each measurement method & 3\\ 
Factor 3 & Covariance between $X$ and $Y$ & 2 \\ 
Factor 4 & Number of agreements & 2 \\ 
Factor 5 & Exclusion zone & 2 \\ 
Factor 6 & Number of subjects & 2 \\ 
Factor 7 & Methods & 3 / 4 \\ 
\hline
\end{tabular}
\end{center}
\end{table}

\begin{table}
\begin{center}
\caption{Mean patterns in Factor 1: Label  $\circ$ indicates the pattern of agreement between $\mu_{X}$ and $\mu_{Y}$ 2 times, $\times$ indicates the pattern of not agreement between $\mu_{X}$ and $\mu_{Y}$ 2 times}  \label{T2}
\begin{tabular}{lrrrrclrrrrc} 
\hline
Pattern No. &$\mu_{X1}$ & $\mu_{X2}$ & $\mu_{Y1}$ & $\mu_{Y2}$ & Label & Pattern No. &$\mu_{X1}$ & $\mu_{X2}$ & $\mu_{Y1}$ & $\mu_{Y2}$ & Label\\  \hline
Pattern1 & -1.5 & -1.5 & 1.5 & 1.5 & $\times$ & Pattern16 & 0.5 & 0.5 & -0.5 & -0.5 & $\times$\\ 
Pattern2 & -0.5 & -0.5 & 0.5 & 0.5 & $\times$ & Pattern17 & -0.5 & -1.5 & -0.5 & -1.5 & $\circ$\\ 
Pattern3 & -1.5 & 1.5 & 1.5 & 1.5 & $\times$ & Pattern18 & 0.5 & -1.5 & -0.5 & -1.5 & $\times$\\ 
Pattern4 & 0.5 & -0.5 & 0.5 & 0.5 & $\times$ & Pattern19 & -0.5 & 1.5 & -0.5 & -1.5 & $\times$\\ 
Pattern5 & 1.5 & 1.5 & 1.5 & 1.5 & $\circ$ & Pattern20 & 0.5 & 1.5 & -0.5 & -1.5 & $\times$\\ 
Pattern6 & 0.5 & 0.5 & 0.5 & 0.5 & $\circ$ & Pattern21 & -1.5 & -1.5 & -1.5 & 1.5 & $\times$\\ 
Pattern7 & -0.5 & -1.5 & 0.5 & 1.5 & $\times$ & Pattern22 & -0.5 & -0.5 & -0.5 & 0.5 & $\times$\\
Pattern8 & 0.5 & -1.5 & 0.5 & 1.5 & $\times$ & Pattern23 & -1.5 & 1.5 & -1.5 & 1.5 & $\circ$\\
Pattern9 & -0.5 & 1.5 & 0.5 & 1.5 & $\times$ & Pattern24 & 0.5 & -0.5 & -0.5 & 0.5 & $\times$\\
Pattern10 & 0.5 & 1.5 & 0.5 & 1.5 & $\circ$ & Pattern25 & 1.5 & 1.5 & -1.5 & 1.5 & $\times$\\
Pattern11 & -1.5 & -1.5 & -1.5 & -1.5 & $\circ$ & Pattern26 & 0.5 & 0.5 & -0.5 & 0.5 & $\times$\\
Pattern12 & -0.5 & -0.5 & -0.5 & -0.5 & $\circ$ & Pattern27 & -0.5 & -1.5 & -0.5 & 1.5 & $\times$\\
Pattern13 & -1.5 & 1.5 & -1.5 & -1.5 & $\times$ & Pattern28 & 0.5 & -1.5 & -0.5 & 1.5 & $\times$\\
Pattern14 & 0.5 & -0.5 & -0.5 & -0.5 & $\times$ & Pattern29 & -0.5 & 1.5 & -0.5 & 1.5 & $\circ$\\
Pattern15 & 1.5 & 1.5 & -1.5 & -1.5 & $\times$ & Pattern30 & 0.5 & 1.5 & -0.5 & 1.5 & $\times$\\
\hline
\end{tabular}
\end{center}
\end{table}

\subsection{Simulation results}
\subsubsection{Difference between the true value and the estimation of each concordance rate method}

\begin{figure}[htbp]
\begin{center}
\includegraphics[scale=0.55]{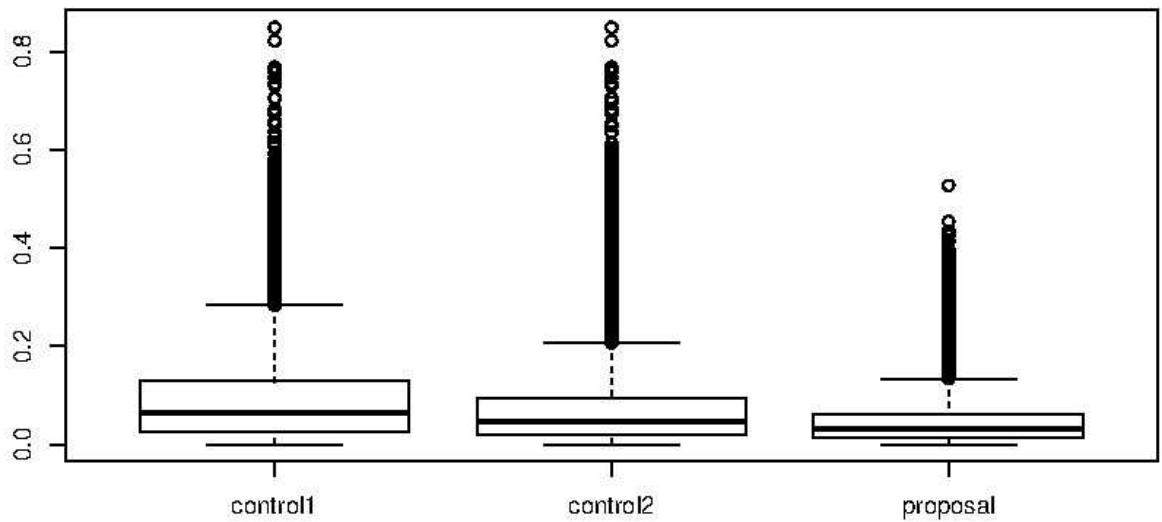}
\caption{Result of the simulation}
\label{fig.1}
\end{center}
\end{figure}

In all simulation results, the proposed approach was closer to the true value than control methods. Figure \ref{fig.1} showed the result of this simulation. We also showed median, the first quartile and the third quartile by each factor in tables. Medians of the proposal method was smaller, and the interquantile range was narrower than than the control1 and control2 in all factors. These results indicate that the variation of the proposed method was smaller than two control concordance methods. The estimation of the proposal was stable.\
Table \ref{TF1} are the results par each pattern of the mean. In Pattern3, 13, 21 and 25, the bias of control1 tended to be large.
These patterns are the situation that the all absolute values of means of $X$ and $Y$ are 1.5 and the direction of trends disagree two times. 
compared to the control methods, the proposed method was stable in all patterns. 
Table \ref{TF2} is the results of the covariance of the difference values within each measurement method and table \ref{TF3} is the results of the covariance between $X$ ans $Y$. The results of all methods were almost same in terms of both covariances. The proposed method was more stable than the conventional methods in both factors.
The proposed method resulted more closely to the true values in both $m=1$ and $m=2$ than the control methods (Table \ref{TF4}). It means that the proposal evaluated more properly in all number of agreement in the case of $T=2$.
Regarding the exclusion zone, the concordance rates was slightly higher in larger size of the exclusion zone (Table \ref{TF5}).  
The concordance rates in all methods were smaller in the larger number of subjects (Table \ref{TF6}).

\subsubsection{Diagnosability of the estimation of each concordance method}
\begin{figure}[h!!!]
\begin{center}
\includegraphics[scale=0.5]{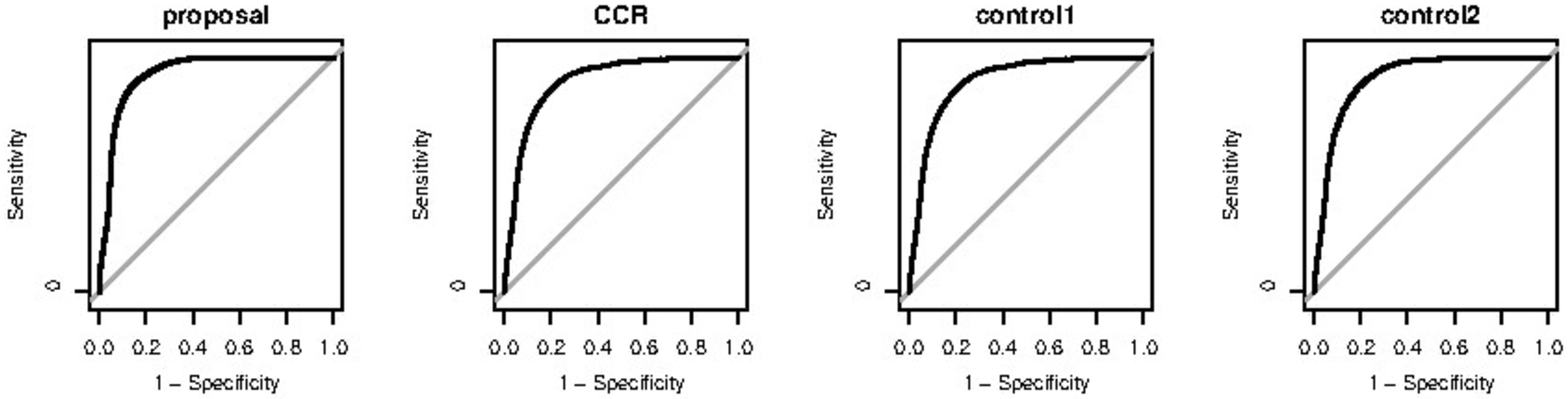}
\caption{ROC curves of proposal method, CCR, control1 and control2 for the simulation}
\label{fig.8.1}
\end{center}
\end{figure}

To compare the diagnosability of the proposed method with that of the conventional methods, we described the ROC curves of the proposal, CCR, control1 and control2 in Figure \ref{fig.8.1} and calculated their AUC in Table \ref{T3}. Seeing from the results of AUC, the  proposed method was better than the conventional methods. In other words, it showed that the diagnostic capability of the proposed method was superior to the conventional methods.

\begin{table}
\begin{center}
\caption{The result of the simulation for Factor1: Means.}\label{TF1}
\begin{tabular}{l|lll} 
\hline
Pattern No. & control1 & control2 & proposal\\ \hline 
Pattern1 & 0.028 (0.011, 0.059) & 0.028 (0.011, 0.060) & 0.018 (0.007, 0.042) \\
Pattern2 & 0.076 (0.036, 0.137) & 0.079 (0.036, 0.141) & 0.047 (0.022, 0.086) \\
Pattern3 & 0.158 (0.131, 0.191) & 0.038 (0.019, 0.069) & 0.025 (0.012, 0.043)\\
Pattern4 & 0.066 (0.030, 0.117) & 0.071 (0.034, 0.123) & 0.046 (0.021, 0.080)\\
Pattern5 & 0.023 (0.010, 0.057) & 0.023 (0.011, 0.057) & 0.015 (0.006, 0.037)\\
Pattern6 & 0.072 (0.034, 0.124) & 0.074 (0.036, 0.127) & 0.043 (0.019, 0.079)\\
Pattern7 & 0.048 (0.022, 0.093) & 0.053 (0.024, 0.096) & 0.033 (0.015, 0.068)\\
Pattern8 & 0.070 (0.035, 0.118) & 0.051 (0.023, 0.093) & 0.033 (0.016, 0.062)\\
Pattern9 & 0.059 (0.029, 0.105) & 0.043 (0.020, 0.086) & 0.028 (0.013, 0.059)\\
Pattern10 & 0.035 (0.016, 0.081) & 0.039 (0.020, 0.085) & 0.025 (0.011, 0.060)\\
Pattern11 & 0.022 (0.010, 0.055) & 0.023 (0.011, 0.056) & 0.014 (0.006, 0.036)\\
Pattern12 & 0.072 (0.035, 0.124) & 0.074 (0.038, 0.126) & 0.042 (0.020, 0.077)\\
Pattern13 & 0.159 (0.132, 0.190) & 0.038 (0.019, 0.069) & 0.025 (0.012, 0.042)\\
Pattern14 & 0.065 (0.029, 0.117) & 0.069 (0.032, 0.127) & 0.045 (0.021, 0.082)\\
Pattern15 & 0.029 (0.011, 0.061) & 0.030 (0.011, 0.062) & 0.018 (0.007, 0.042)\\
Pattern16 & 0.079 (0.038, 0.137) & 0.082 (0.038, 0.141) & 0.047 (0.021, 0.086)\\
Pattern17 & 0.036 (0.016, 0.085) & 0.040 (0.021, 0.086) & 0.026 (0.011, 0.060)\\
Pattern18 & 0.059 (0.029, 0.104) & 0.043 (0.020, 0.090) & 0.030 (0.013, 0.061)\\
Pattern19 & 0.070 (0.034, 0.116) & 0.052 (0.024, 0.095) & 0.034 (0.015, 0.062)\\
Pattern20 & 0.047 (0.021, 0.092) & 0.052 (0.023, 0.092) & 0.033 (0.014, 0.067)\\
Pattern21 & 0.159 (0.132, 0.190) & 0.038 (0.019, 0.069) & 0.024 (0.011, 0.042)\\
Pattern22 & 0.066 (0.031, 0.117) & 0.073 (0.035, 0.127) & 0.046 (0.021, 0.082)\\
Pattern23 & 0.015 (0.005, 0.056) & 0.014 (0.005, 0.056) & 0.009 (0.003, 0.036)\\
Pattern24 & 0.068 (0.030, 0.120) & 0.071 (0.031, 0.125) & 0.045 (0.021, 0.080)\\
Pattern25 & 0.158 (0.130, 0.190) & 0.038 (0.019, 0.069) & 0.025 (0.012, 0.043)\\
Pattern26 & 0.067 (0.030, 0.118) & 0.074 (0.034, 0.128) & 0.046 (0.022, 0.082)\\
Pattern27 & 0.072 (0.034, 0.117) & 0.050 (0.023, 0.094) & 0.034 (0.016, 0.064)\\
Pattern28 & 0.046 (0.020, 0.086) & 0.045 (0.020, 0.090) & 0.031 (0.013, 0.062)\\
Pattern29 & 0.042 (0.017, 0.089) & 0.036 (0.016, 0.084) & 0.022 (0.009, 0.055)\\
Pattern30 & 0.058 (0.028, 0.103) & 0.042 (0.019, 0.086) & 0.029 (0.013, 0.059)\\
\hline
\end{tabular}\\
\end{center}
$\quad \quad \quad \quad \quad \quad \quad$ median(first quartile, third quartile)
\end{table}

\begin{table}
\begin{center}
\caption{The result of the simulation for Factor2: Covariance of the difference values within each measurement method.}\label{TF2}
\begin{tabular}{l|lll} 
\hline
 & control1 & control2 & proposal \\ \hline 
 ${\rho} = 0$ & 0.061 (0.022, 0.123) & 0.043 (0.017, 0.091) & 0.028 (0.011, 0.058)\\
 ${\rho} = 1/3$ & 0.063 (0.022, 0.124) & 0.045 (0.019, 0.091) & 0.029 (0.012, 0.060)\\
  ${\rho} = 2/3$ & 0.069 (0.029, 0.137) & 0.052 (0.025, 0.102) & 0.033 (0.014, 0.065)\\
\hline
\end{tabular}
\end{center}
$\quad \quad \quad \quad \quad \quad \quad$ median(first quartile, third quartile)
\end{table}

\begin{table}
\begin{center}
\caption{The result of the simulation for Factor3:  Covariance between $X$ and $Y$}\label{TF3}
\begin{tabular}{l|lll} 
\hline
 & control1 & control2 & proposal\\ \hline 
${\rho}_{XY} = 0$ & 0.065 (0.025, 0.127) & 0.048 (0.02, 0.094) & 0.031 (0.013, 0.062)\\
${\rho}_{XY} = 1/3$ & 0.064 (0.024, 0.130) & 0.046 (0.019, 0.095) & 0.029 (0.012, 0.060)\\
\hline
\end{tabular}
\end{center}
$\quad \quad \quad \quad \quad \quad \quad$ median(first quartile, third quartile)
\end{table}

\begin{table}
\begin{center}
\caption{The result of the simulation for Factor4:  Number of agreements.}\label{TF4}
\begin{tabular}{l|lll} 
\hline
 & control1 & control2 & proposal\\ \hline 
$m = 1$ & 0.059 (0.021, 0.122) & 0.042 (0.017, 0.087)  & 0.027 (0.011, 0.056)\\
$m = 2$ & 0.070 (0.029, 0.135) & 0.052 (0.023, 0.102) & 0.034 (0.014, 0.066)\\
\hline
\end{tabular}
\end{center}
$\quad \quad \quad \quad \quad \quad \quad$ median(first quartile, third quartile)
\end{table}

\begin{table}
\begin{center}
\caption{The result of the simulation for Factor5: Exclusion zone. }\label{TF5}
\begin{tabular}{l|lll} 
\hline
 & control1 & control2 & proposal\\ \hline 
$a = 0.5$ & 0.058 (0.023, 0.116) & 0.043 (0.019, 0.085) & 0.028 (0.012, 0.055)\\
$a = 1.0$ & 0.072 (0.026, 0.141) & 0.053 (0.021, 0.106) & 0.032 (0.013, 0.067)\\
\hline
\end{tabular}
\end{center}
$\quad \quad \quad \quad \quad \quad \quad$ median(first quartile, third quartile)
\end{table}

\begin{table}
\begin{center}
\caption{The result of the simulation for Factor6: Number of subjects. }\label{TF6}
\begin{tabular}{l|lll} 
\hline
 & control1 & control2 & proposal\\ \hline 
$n = 15$ & 0.077 (0.029, 0.145) & 0.061 (0.025, 0.117) & 0.039 (0.016, 0.078)\\
$n = 40$ & 0.055 (0.022, 0.111) & 0.038 (0.016, 0.074) & 0.024 (0.010, 0.047)\\
\hline
\end{tabular}
\end{center}
$\quad \quad \quad \quad \quad \quad \quad$ median(first quartile, third quartile)

\begin{center}
\caption{AUC of proposal method, CCR, control1 and control2 in the simulation}\label{T3}
\begin{tabular}{l|llll} 
\hline
 & proposal & CCR & control1 & control2\\ \hline
AUC & 0.930 & 0.898 & 0.898 & 0.909\\ 
\hline
\end{tabular}
\label{table3}
\end{center}
\end{table}

\newpage

\section{Real Example}
\label{sec:5}

In this section, we show the usefulness of the proposed concordance rate by the diagnosability through a real example. 

\begin{figure}[htbp]
\begin{center}
\includegraphics[scale=0.6]{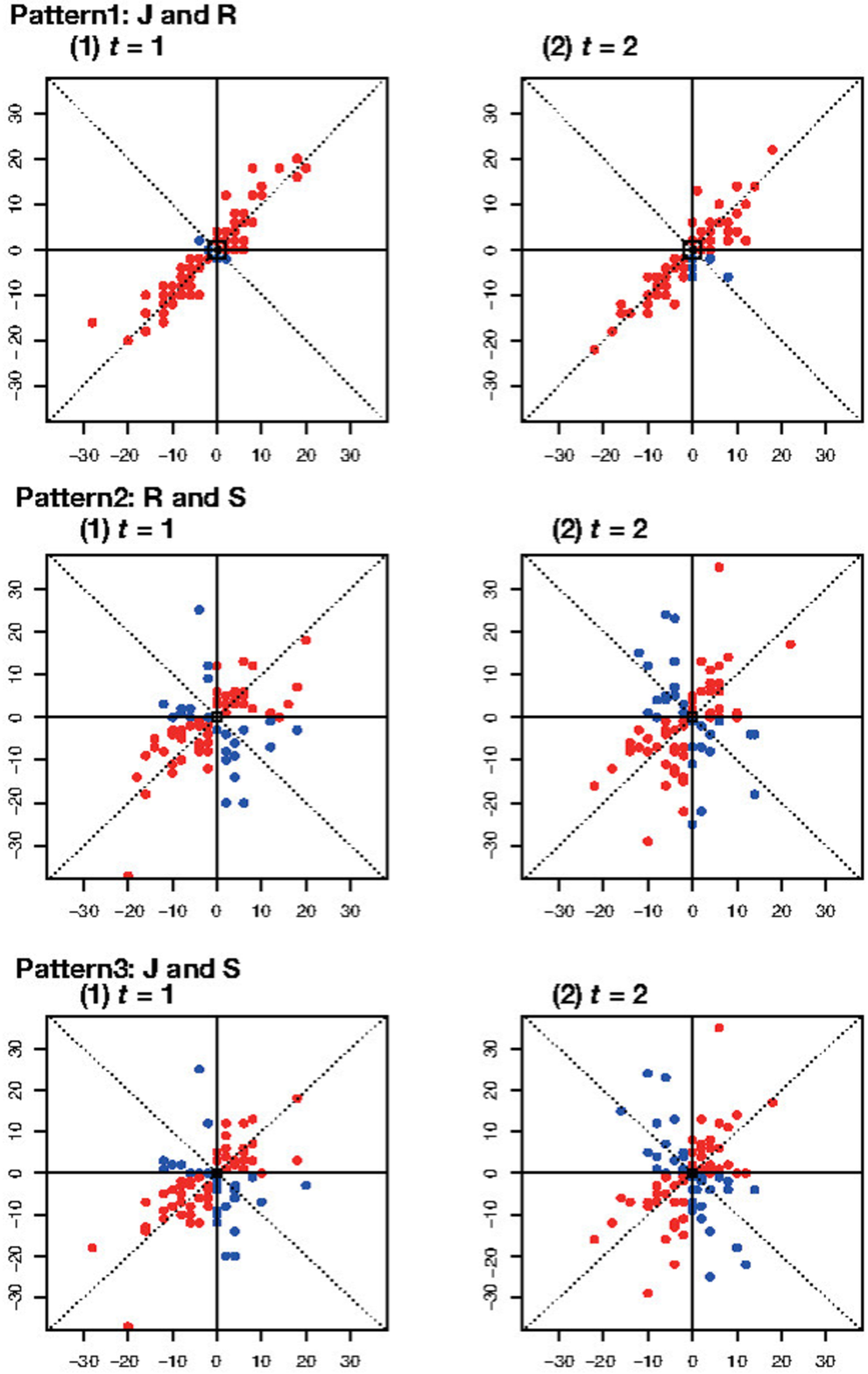}
\caption{Four-quadrant plots with real example data. Pattern1: J(observer1) and R(observer2), Pattern2: R and S(automatic machine), and Pattern3: J and S.}
\label{fig.9}
\end{center}
\end{figure}

We applied the proposed methods to the blood pressure data of package {\choosefont{pcr}MethComp} in 
{\choosefont{pcr} R software} (Carstensen {\it{et al}}., 2020). The data (Altman and Bland, 1991; Bland and Altman, 1999) comprise the blood pressure measurement for 85 subjects based on 3 types of data: data named as J and R were measured by a gold standard conducted by 2 different human observers, and S was measured by an automatic machine as the experimental method. The study was performed at three time points for each subject.\
The four-quadrant plots generated from the real data are presented in Figure \ref{fig.9}. 
Comparing 2 of the 3 measurement results to one another, there are 3 pairs: J(observer1) and R(observer2), R and S(auto machine), and J and S. Each pattern has 2 plots, (1) $t=1$ and (2) $t=2$. We calculated the concordance rate with the proposed method, CCR, control1 and control2 as described in Section 4. The concordance rate was in the 2 cases when the trend of change agreed once in two time points ($m = 1$) and twice all time points ($m = 2$). 
${\rm Ez}(a)$ was set as 10 percent quantile point in each pair (e.g. Critchley {\it{et al}}., 2010). \

As the assessment of the methods, we compared the diagnostic feasibility of the proposal and the conventional methods of CCR, control1 and control2. Specifically, each 10 subjects out of 85 were randomly selected $100$ times, and the concordance rates was obtained by the four methods in the only case of $m=2$ in each pair.
Based on the results, AUC of the proposal, CCR, control1 and  control2 were calculated, and ROC curves of the proposal and CCR were drawn to estimate the diagnosability.
%

%

Each pattern of the four-quadrant plots in Figure \ref{fig.9} shows the characteristics of the real example. The Data of J and R in Pattern 1 have many red points which show "agreement" of the trend between two data and most of these points lie close to the $45^\circ$ line, because this tendency naturally derives from the same established measurement method. On the other hand, data of S, the experimental measurement, is collected in the different way, thus the plots of Pattern2 and Pattern3 have more blue dots as "disagreement" than the plots of pattern1, and the data are distributed with variation.
Then, data of pattern 1 is attached "agreement" label, and data of both pattern2 and pattern 3 as "disagreement" label. 
For the evaluation, $10$ subjects out of 85 are randomly selected and calculated by proposal, CCR, control1, and control2 inall three patterns.
The procedure was iterated $1000$ times and the diagnostic performances of each method are evaluated.

\begin{figure}[htbp]
\begin{center}
\includegraphics[scale=0.7]{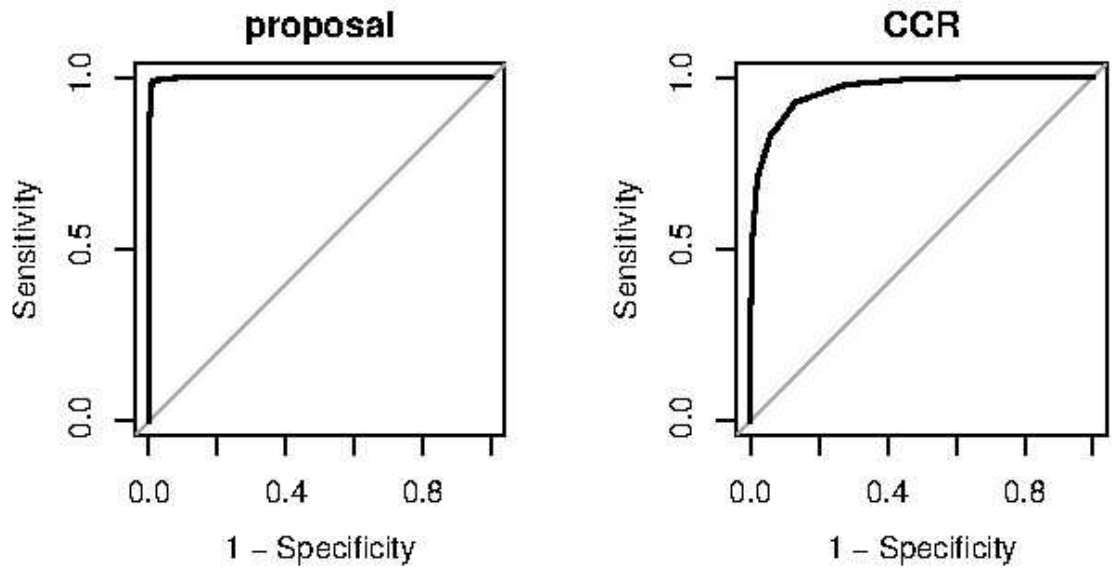}
\caption{ROC of proposal and CCR}
\label{fig.10}
\end{center}
\end{figure}

AUC of the proposal methods, CCR, control1 and control2 shows in Table \ref{Table4}. Each concordance rate estimated with high accuracy in $m=2$ of the example data. The proposal was better than CCR, control1 and control2. 
As for ROC curves in Figure \ref{fig.10}, the plot of the proposed method drew a curve with almost right angle, while the curve was more moderate in the ROC of CCR.
The AUC and ROC curves indicate that the proposed approach has more accuracy than the conventional concordance rates.


\begin{table}[htbp]
\begin{center}
\caption{AUC of CCR, control1, control2 and proposal in a real example}\label{R3}
\begin{tabular}{llll} 
\hline
proposal &CCR & control1 & control2   \\ \hline
$0.999$ & $0.964$ & $0.964$ & $0.965$  \\
\hline
\label{Table4}
\end{tabular}
\end{center}
\end{table}

\section{Discussion} 
\label{sec:6}

The conventional concordance rate for a four-quadrant plot is one of the methods for evaluating the equivalence between a new testing method and a standard measurement method. In many clinical practice situations, these values are observed repeatedly for the same subjects. However, the conventional concordance rate for the four-quadrant plot does not consider individual subjects when evaluating the trend of measurement values between two clinical testing methods being compared. Therefore, we proposed a new concordance rate based on normal distribution that is calculated using the difference values in each measurement technique depending on the number of agreements. The minimum number of agreements to evaluate the equivalence named hyper parameter can be set according to the total number of time points in the data and the clinical point of view. 

In most factors set in the simulation, the proposed concordance rate was mostly closer to the true value than the conventional methods.
In addition to that, the diagnosability of the estimation of the proposed method
was superior to both the existing concordance method and its applied control methods from the results of numerical simulations.
%
In addition, through the real example using sbp data,
we showed the superiority of the proposed method for the diagnosability by these AUC values.
We also provided only the results of the numerical simulations and a real example for the case of time point $T = 2$ in this study; however, this proposed concordance rate can be calculated as a case of any $T$.

Here we mention the assumptions of the proposed method and its comparison with existing statistical methods.
In the proposed method, we assumed that these data are distributed from multivariate normal distribution.
In practical situation, concordance rate is used with Bland-Altman analysis to
evaluate the equivalence of two measurement methods.
Bland-Altman analysis assumed to be distributed from normal distribution (e.g. Bland and Altman, 2007; Bartko, 1976; Zou, 2013).
Therefore, the assumption of the proposed method is consistent with that of Bland-Altman analysis.
Next, Goodman and Kruskal's gamma (Goodman and Kruskal, 1963) is similar to the concordance rate, although the range is different. The gamma statistic does not consider the exclusion zone and, in the practical situation of clinical trials, concordance rate is usually used with Bland-Altman analysis.

Finally, We further discuss the four points of future work of this study. First, for the values of the proposed concordance rate, there are no absolute criteria, similar to the conventional concordance rate. Although various criteria have been proposed, there are no common acceptable criteria for the conventional concordance rate (e.g., Saugel {\it{et al}}., 2015). Therefore, it is difficult to determine the result as good, acceptable, or poor. Secondly, the results of the proposed concordance rate may also face the problem at time intervals between the measurement values, similar to the conventional concordance rate (e.g., Saugel {\it{et al}}., 2015). Thirdly, we have to determine the parameters of the exclusion zone (e.g., Critchley {\it{et al}}., 2011).
Forthly, in the proposed method, we introduced hyper parameter $m$, which allows us to arrive at a flexible interpretation of the results. While the Bland-Altman analysis was sometimes used in confirmatory clinical trials based on the statistical inference (e.g., Asamoto {\it{et al}}., 2017), our proposed concordance rate for the four-quadrant plot has not been established yet in this regard. The estimation of the confidence interval will be needed.

In this study, we found that the conventional concordance rate was not so proper indicator in repeated measurements, while the proposed concordance rate could enhance the accuracy by calculating depending on the number of agreement. As the proposed concordance rate provides the trending agreement from various perspectives, this new method is expected to contribute to clinical decisions as an exploratory analysis. Further consideration is thus required from these points of view.

\

\noindent {\bf{Conflict of Interest}}

\noindent {\it{The authors have declared no conflict of interest. }}

\bibliographystyle{unsrt}  


\end{document}